# On the Cost of Cyber Security in Smart Business

Igor Ivkic, Stephan Wolfauer, Thomas Oberhofer, Markus G. Tauber
University of Applied Sciences Burgenland
Eisenstadt, Austria
e-mail: {igor.ivkic,stephan.wolfauer,thomas.oberhofer,markus.tauber}@fh-burgenland.at

*Abstract*—**In a world, as complex and constantly changing as ours cloud computing is a driving force for shaping the IT landscape and changing the way we do business. Current trends show a world of people, things and services all digitally interconnected via the Internet of Things (IoT). This applies in particular to an industrial environment where smart devices and intelligent services pave the way for smart factories and smart businesses. This paper investigates in a use case driven study the potential of making use of smart devices to enable direct, automated and voice-controlled smart businesses. Furthermore, the paper presents an initial investigation on methodologies for measuring costs of cyber security controls for cloud services.**

***Keywords—cloud-services; security controls; cost estimation***

## I. Introduction

The internet boom in the late 90s opened the gates to a new era of the information age [1]. In this era software applications were shaped in world-wide connectivity, look and feel and the time to market [2] decreased significantly. On top of the world wide web, mobile devices along with app stores expanded the boundaries of personal computers (PC). Instead of being bound to the stationary PCs at home, these devices enabled users to operate globally. In addition to that the app stores invited everybody with a computer to professionally develop and sell apps. The latest "game changer" in information technology (IT) transformed the landscape of infrastructure by declaring on premise installations as obsolete in a nearby future. Cloud computing, as Armbrust et al. [3] stated, offers a diverse variety of services, allocates the necessary infrastructure resources and charges in a pay-as-you-go manner. As a result, many of today's big companies were built on good ideas implemented as apps or services and operated in the cloud.

Besides cloud computing the IoT and Industry 4.0 were among the latest IT trends [4]. IoT in combination with cloud computing promises to turn our computers and automated machines into Cyber Physical Systems (CPS) [5] which leads to Industry 4.0 [4]. In an Industry 4.0 "smart factory" [6], CPS are communicating with other CPS and with human beings in real time over the IoT. Even though these two trends are very promising there are a lot of challenges to overcome before an Industry 4.0 becomes reality. One of them is to guarantee a secure communication in CPS [7] and to measure the effort it takes (cost) in order to do so. Another challenge is the simple lack of prototypes to demonstrate a functioning IoT based, secure and cost-efficient "smart business".

In this paper, we evaluate whether a smart device could be introduced to a company to communicate via a cloud services and to automate manually performed tasks. Building on this we present a business-model and its architectural design in a use case driven study including security considerations. In a final step, we apply security controls suggested by ISO 27017 [8] in combination with Six Sigma [9] to first, eliminate security risks in the presented use cases and to second, measure additional expenses resulting from them. Our contribution in this paper towards applying an established security standard and a quality management method in a use case driven study is twofold:

- firstly, we derive its business need and the required technical design including security considerations and
- secondly, we present an initial approach to eliminate security risks and measure the effort it takes to do so.

The rest of the paper is organized as follows: Section II summarizes the related work in the field, followed by the presentation of a use case study in Section III. Finally, in Section IV we introduce a high-level process flow based on Six Sigma for identifying, categorizing, analyzing, eliminating security risks and measuring the resulting costs.

## II. Related Work

Guaranteeing security and measuring the resulting costs of cloud services faces considerable difficulties. Related work in [10] and [11] has developed research results about improving the service quality using Six Sigma. In contrast to that the work in [12] presented a management process for "Security Policy Management" within the Six Sigma framework. In [13] Chen & Sion present solutions of how to protect data and guarantee security in the public cloud. However, none of these works presented solutions of how to apply ISO 27017 controls to secure cloud services and how to measure the effort it takes to guarantee security of a cloud service. This paper presents an approach for applying an established standard to secure a cloud service and to use an established quality management method to measure resulting costs. In this regard, this papers approach offers a method for guaranteeing security of cloud services and quantifying the implementation expenses.

## III. Business-Model & Use Case

Amazon Echo and the Amazon Web Services (AWS) meet all requirements needed in the following business-model and Use Cases I and II. Figure 1 shows the basic concept of Amazon Echo communicating with the AWS via Amazon Voice Service (AVS) including the processing of the user commands via Alexa service, the Spoken Language Understanding (SLU) and the Amazon Skill Kit (ASK) [14]:

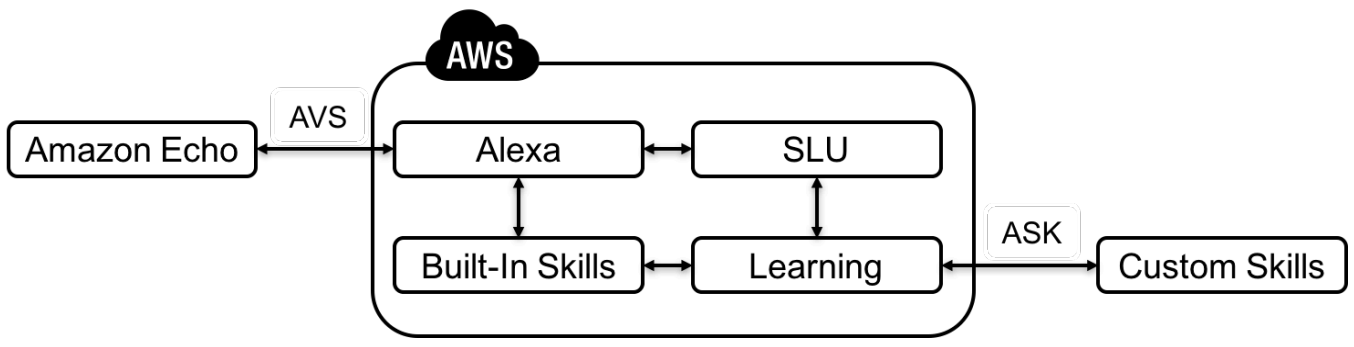

Figure 1. Amazon Echo & AWS Technology. (adapted from Soper, 2016)

We use Amazon Echo and AWS to develop a smart-business case for a fictive company with branches in two cities (City-A and City-B). The company's core business is delivering packages to customers. At first, an ordered package is delivered to one of the two branches and then transported by delivery trucks to the customers. Usually, the truck is loaded in the morning in City-A before the driver departs to City-B delivering packets on the way. Typically, on the following day, the same truck departs in City-B and heads back to City-A supplying customers on its way, and so on. The key to success for the business model is the organization and direct communication between the two branches and their delivery truck drivers. To save time and costs, voice controlled messaging (active and passive) is considered for automated routine jobs via Amazon Echo. Three deployed devices use AWS for computation (i.e.: convert natural language in computer understandable language, vice versa) and as a dispatching service (i.e.: sending voice mails). For the use case (Figure 2) we also consider ISO 27017 and Six Sigma.

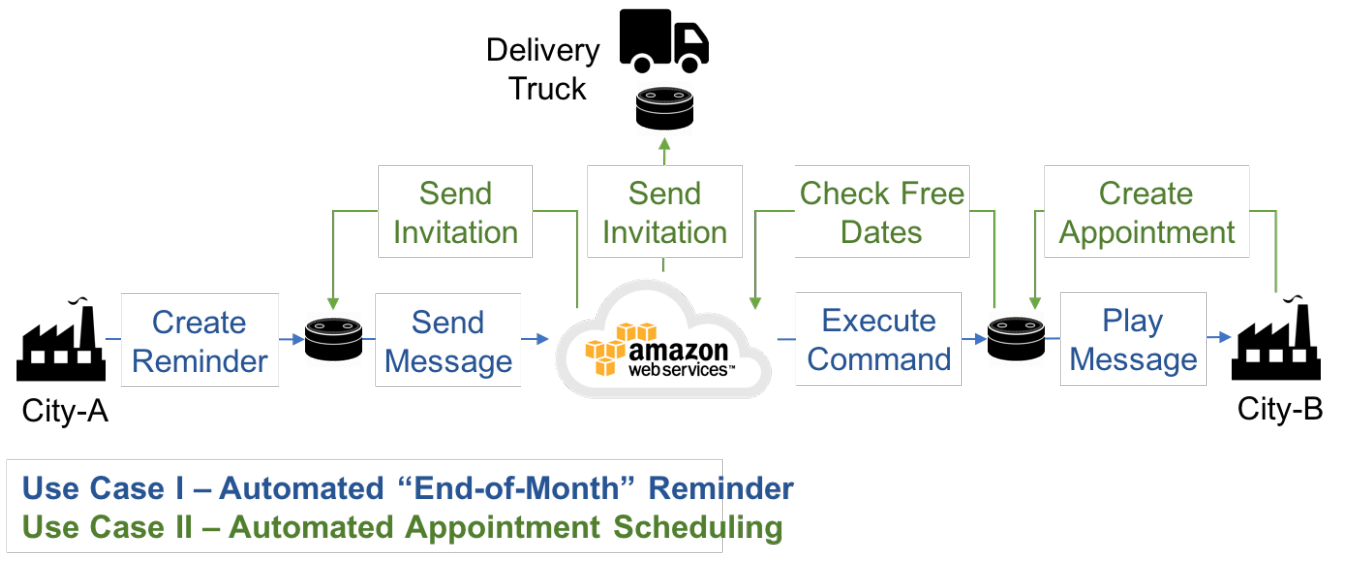

Figure 2. Use Case – Architecture, based on use case description below

## A. *Use Case I*

The finance manager in City-A is planning to improve some accounting tasks, especially the accounting process at the end of a month. Usually, each employee is committed to daily time recording which must be released at the end of the month. Only when the time recordings of an employee have been released the employees of the finance department can start with the accounting process. Unfortunately, it repeatedly happens, that employees forget to release their time recordings in time. This mistake leads to a delay of accounting which means increased costs of administration and effort to finally finish the accounting process. The following sequence diagram shows the Use Case I as described in the next paragraphs:

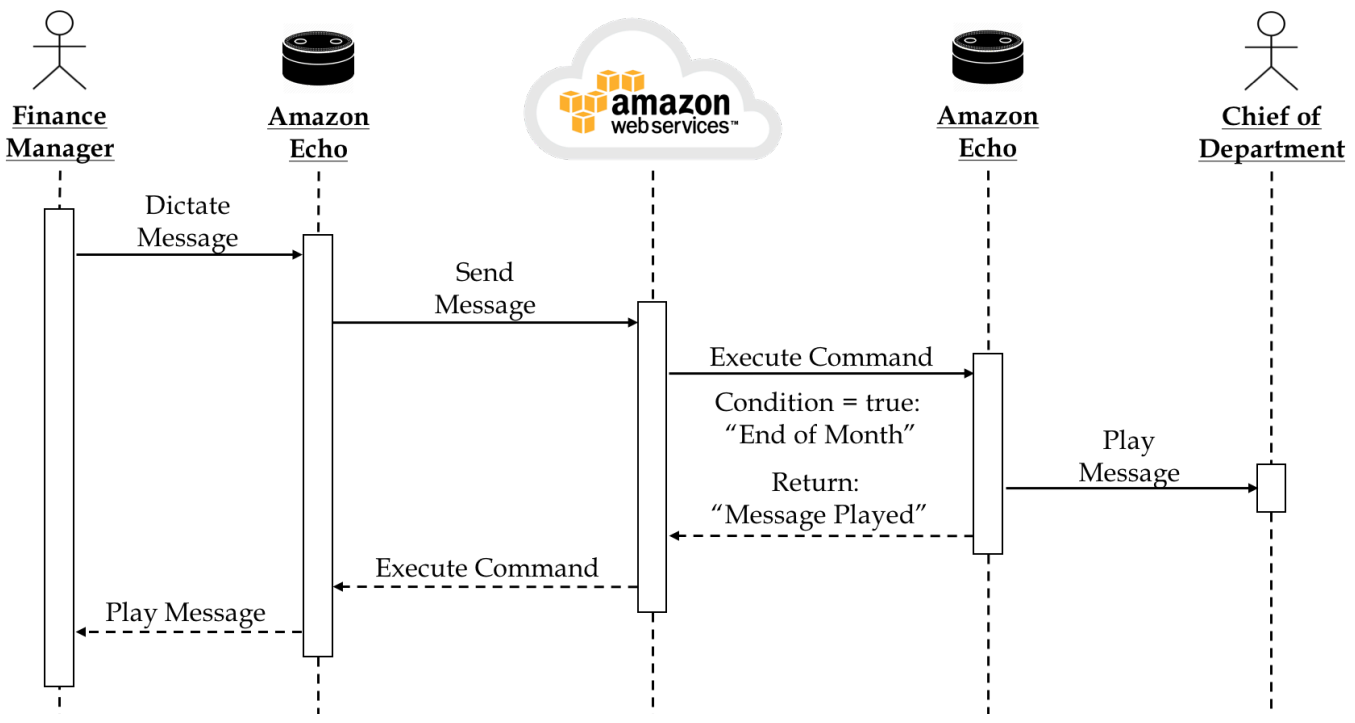

Figure 3. Use Case I – Automated "End-of-Month" Reminder.

So, to fix this issue once and for all the finance manager dictates a voice message to Alexa with the condition that his message should be played to the Chief of Department in City-B at the end of each month. The dictated message is meant to remember the colleague from the other branch to remind the employees to release their time recordings in time.

Use Case I in Figure 3 shows the entire "Reminder Process" starting with the message dictated by the Finance Manager to the message being played to the Chief of Department in City-B. The Use Case I does not show, that the process above is created one time and then repeatedly executed at the end of every month. All the commands necessary to understand and execute the Finance Managers requests are provided by the AWS.

## B. *Use Case II*

The Chief of Department in City-B plans to invite the Finance Manager and a Truck Driver to talk about a customer complaint. Instead of using his email client and wasting time on searching for a date where all three attendees were free, he decides to ask Alexa for help. He uses the key word "Alexa Calendar" to organize an appointment on the next possible date for all attendees. Alexa sends the request to the AWS which combs through all calendars until a date is found and finally sends an invitation to all attendees. The following sequence diagram shows the described Use Case II:

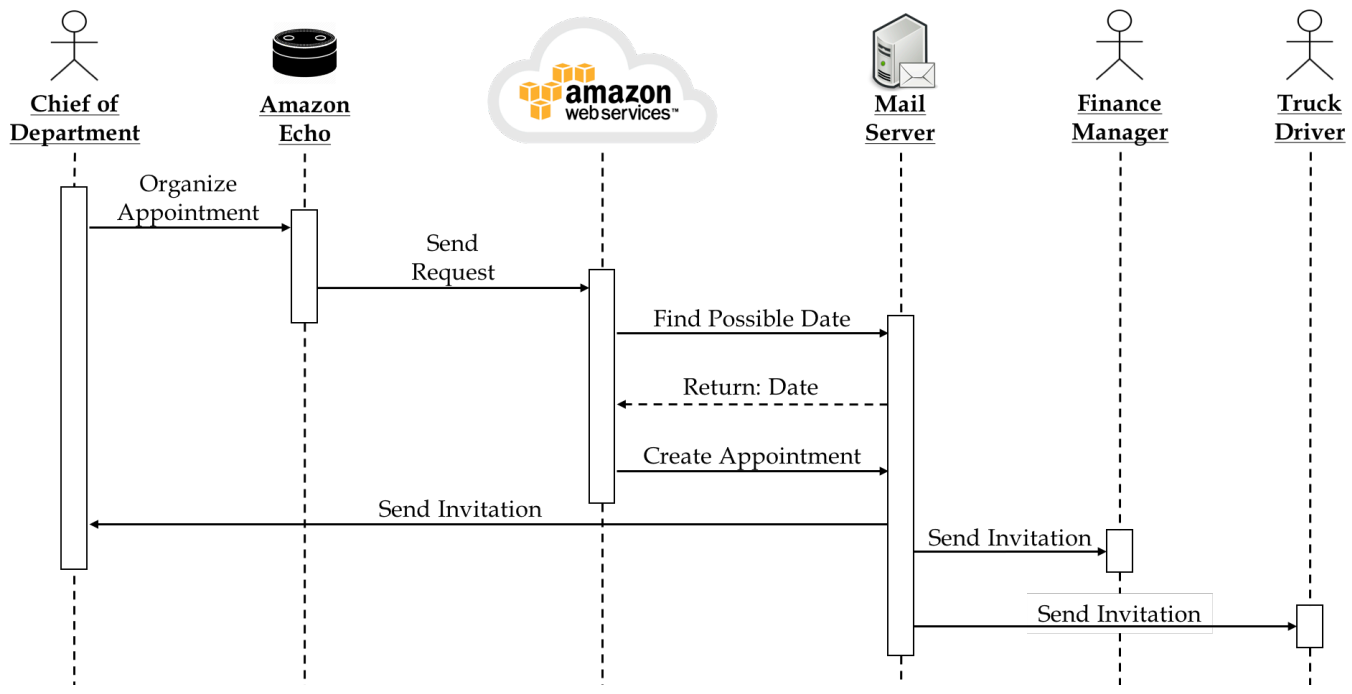

Figure 4. Use Case II – Automated Appointment Scheduling.

Figure 4 shows how the additional features of the business-model can be used to combine the AWS with i.e.: a mail server. In this scenario manually performed tasks like "creating a new appointment", "searching for free dates in calendars" and "sending an invitation to a meeting" become obsolete. It is another nice example of how simple it is to use cloud services as a building block system to improve working routines or even entire processes.

# IV. Evaluation

After presenting the business-model architecture and the two use cases in the previous chapter, the evaluation in the following is three-fold. First, the Open Web Application Security Project (OWASP) top ten cloud security risks are categorized by their importance to the presented use cases (Section A). Based on that ranking, the ISO 27017 is used then to eliminate the top three risks (Section B). In this step, it is demonstrated how the security controls as proposed by the ISO standard are applied to eliminate the risks in the use cases. Finally, an adapted version of the Six Sigma method, is proposed to identify, categorize, analyze, eliminate and measure security risks (Section C).

### *A. Cloud Security Risks*

The OWASP is a non-profit organization operating world-wide with the main goal to improve the security of software. This organization is on a mission to point out security risks to individuals and organizations to create a base for informed decision making about security. In order to achieve this goal, the OWASP community started many projects focusing on different issues. One of these projects is the "OWASP Cloud – 10 Project" with the goal to maintain a list of top 10 security risks faced with cloud computing. The following list shows the top 10 cloud computing risks according to OWASP (detailed description of risks in [15]):

- R1 – Accountability and Data Ownership
- R2 – User Identity Federation
- R3 – Regulatory Compliance
- R4 – Business Continuity and Resiliency
- R5 – User Privacy and Secondary Usage of Data
- R6 – Service and Data Integration
- R7 – Multi Tenancy and Physical Security
- R8 – Incidence Analysis and Forensic Support
- R9 – Infrastructure Security
- R10 – Non-Production Environment Exposure

Generally, all ten risks should be analyzed and eliminated to guarantee maximum security for a cloud service. Unfortunately, that would be beyond the scope of this paper. Therefore, a risk analysis [16] [17] [18] was carried out to categorize the top ten risks by their relevance to the use cases and their severity of consequences. Figure 5 shows a risk matrix including the results of the risk analysis:

| Relevance | Severity of Consequences | | | | |
|---|---|---|---|---|---|
| | Very Low | Low | Medium | High | Very High |
| Very High | | | | R9 | R6 |
| High | | | R10 | | R4 |
| Medium | | | R3, R7, R8 | | |
| Low | | | | | |
| Very Low | R2 | | | R1, R5 | |

Figure 5. Risk Analysis of OWASP Top 10.

In the presented use cases the smart devices (Amazon Echo) and cloud services (AWS) are mainly used to automate tasks which are usually performed manually. So, the cloud service uses input (voice commands) to either inform/remind other users or create appointments automatically. Because all these tasks could be performed manually by the user without creating additional risks (caused by cloud computing) R1 & R5 were categorized as "very low" (relevance) / "high" (severity of consequences). R2 can be left out of the discussion, because only one cloud provider (AWS) was used in the use cases. The next three risks (R3, R7 and R8) would have been significant if almost all cloud providers did not enable the option to choose the location of your data. But, they are still very serious risks which should not be handled carelessly. Just as much prudence is needed with R10, because security is as much important to productivity systems as it is to development environments. Although, in the presented use cases, there is no need to constantly redevelop the services. So, R10 is an important risk, but not the most important one in the given use cases.

The most important issues in Use Case I and II are R4, R6 and R9. The highest ranked issue is R6, because all information transferred from smart device to smart device (Amazon Echo to Amazon Echo, end-to-end) via cloud services (AWS) must be encrypted. Next on the list is infrastructure security (R9) which means that the smart devices need to be physically secured from theft and/or misuse. Finally, the business must go on no matter what. This does not mean that the cloud services need extra treatment. It means that it is important to have backup devices in case of dysfunction to keep the users (and therefore the business) non-stop available.

### *B. ISO 27017*

The ISO 27017 is a code of practice for information security controls for cloud services based on the standards ISO 27001 and ISO 27002 [19]. In this regard, ISO 27017 provides guidance on information security aspects of cloud computing, recommending and assisting with the implementation of cloud-specific information security controls [19]. The standard was published in 2015 [19] and provides cloud-based guidance on 37 of the controls in ISO 27002 [20]. In addition to that it addresses the following seven cloud controls [20]:

- WHO is responsible FOR WHAT (customer/provider)
- removal/return of assets on terminated contracts
- protection and separation of the customer's virtual environment
- virtual machine configuration
- administrative operations and procedures associated with the cloud environment
- cloud customer monitoring of activity (in the cloud)
- virtual and cloud network environment alignment

When using ISO 27017 in combination with ISO 27001 both the cloud service provider and the cloud service customer benefit of the standards. For instance, when looking at section A.9.4.1 of ISO 27001 in regard of restricting access, the guidance suggests the following controls [19]:

- The cloud service customer should ensure that access to information in the cloud service can be restricted in accordance with its access control policy and that such restrictions are realized.

- The cloud service provider should provide access controls that allow the cloud service customer to restrict access to its cloud services, its cloud service functions and the cloud service customer data maintained in the service.

The following table shows the suggested changes by ISO 27017 to the established standards ISO 27001 / 27002 [21]:

TABLE I. COMPARISON OF CHANGES IN ISO STANDARDS (ADVISERA, 2015)

| Section | ISO 27001 / ISO 27002 Control Section | Level of Change in ISO 27017 |
|---|---|---|
| S5 | Information Security Policy | MODERATE |
| S6 | Organization of Information Security | MODERATE |
| S7 | Human Resource Security | LOW-MODERATE |
| S8 | Asset Management | LOW-MODERATE |
| S9 | Access Control | HIGH |
| S10 | Cryptography | MODERATE |
| S11 | Physical and Environmental Security | LOW-MODERATE |
| S12 | Operations Security | MODERATE-HIGH |
| S13 | Communications Security | MODERATE-HIGH |
| S14 | System Acquisition, Development and Maintenance | MODERATE |
| S15 | Supplier Relationships | MODERATE-HIGH |
| S16 | Information Security Incident Management | MODERATE |
| S17 | Information Security Aspects of Business Continuity | LOW |
| S18 | Compliance | MODERATE-HIGH |

To eliminate the risks R4, R6 and R9 the following security controls as suggested by ISO 27017 should be applied:

- S17 – Information Security Aspects of Business Continuity (to eliminate R4)
- S10 – Cryptography (to eliminate R6)
- S9 – Access Control (to eliminate R9)

There are two security controls in the ISO standard to eliminate R4. To implement the first, control the continuity of information security must be planned, implemented and reviewed. This control is realized by the proposed Six Sigma process in section C. The second control suggests having sufficient redundancies to satisfy availability requirements. In both use cases the availability of the cloud services is guaranteed by the fault tolerance and high availability of AWS [22]. But, it is also important to have enough hardware reserves in case of smart device failure. So, the delivery trucks and the two offices should each at least have one backup smart device (Amazon Echo) and a process to order a replacement.

To guarantee service and data integration (R6) in cloud services the ISO standard suggests implementing cryptographic controls. These controls take care of secure data transferring and authentication, so that the messages can neither be intercepted nor sent/received by unauthorized users/devices. Especially the control to ensure that the person who uses the smart device is really who she/he claims to be is a major requirement for both use cases. Although, user biometric authentication can be achieved by available products like ArmorVox [23], they go hand in hand with high expenses.

The last risk (R9) can be eliminated by the access controls. As suggested by the ISO standard the smart devices need physical security controls, as well as authentication controls for the voice control functionality. To ensure physical security the devices should be locked with i.e.: Kensington desk mount cables. This control provides security against theft in the offices and/or the delivery trucks. Additionally, it is important to implement user management processes to take care of creation of users and password policies, as well as allocation of access rights and special restrictions for privileged access rights. The process must include regular reviews and updates of access rights to continuously verify and improve access controls. Furthermore, the implemented access controls should take effect each time the smart device is used. In this regard, it is important to prohibit the usage of a smart device (especially with the AWS skill to access user mail and calendar data) without an authorized user and a password. So, before a user is permitted to use the smart device she/he must authenticate herself/himself with a code or a password or i.e.: the finger scan sensor on a mobile phone.

### *C. Six Sigma*

Six Sigma was developed in 1986 by Motorola and is a method providing tools for organizations to improve their business processes. The idea behind Six Sigma was to remove causes of errors when detected before they lead to defects in a product or service. This is accomplished by setting up a management system that identifies errors and provides methods for eliminating them. There are two methodologies used within Six Sigma. First the DMAIC (Define, Measure, Analyze, Improve, and Control) method, which is used for improving existing business processes and second the DMADV (Define, Measure, Analyze, Design, and Verify) method for create new processes and new products or services. There are also many different management tools used within Six Sigma [23]. For the sake of completeness, the DMADV method is mentioned here, but will not be discussed in the proposed Six Sigma process.

In the following proposal, the established and well documented DMAIC method is adapted to create a new process. This process includes steps to identify cloud security risks, categorize them, search for a solution, create an implementation plan and finally measure how much effort it would take to apply the elaborated security controls. The DMAIC methodology perfectly suits the task, because of its problem-solving nature and its process-step structure. Figure 6 shows the proposed high-level process flow of the DMAIC method through its five steps [24].

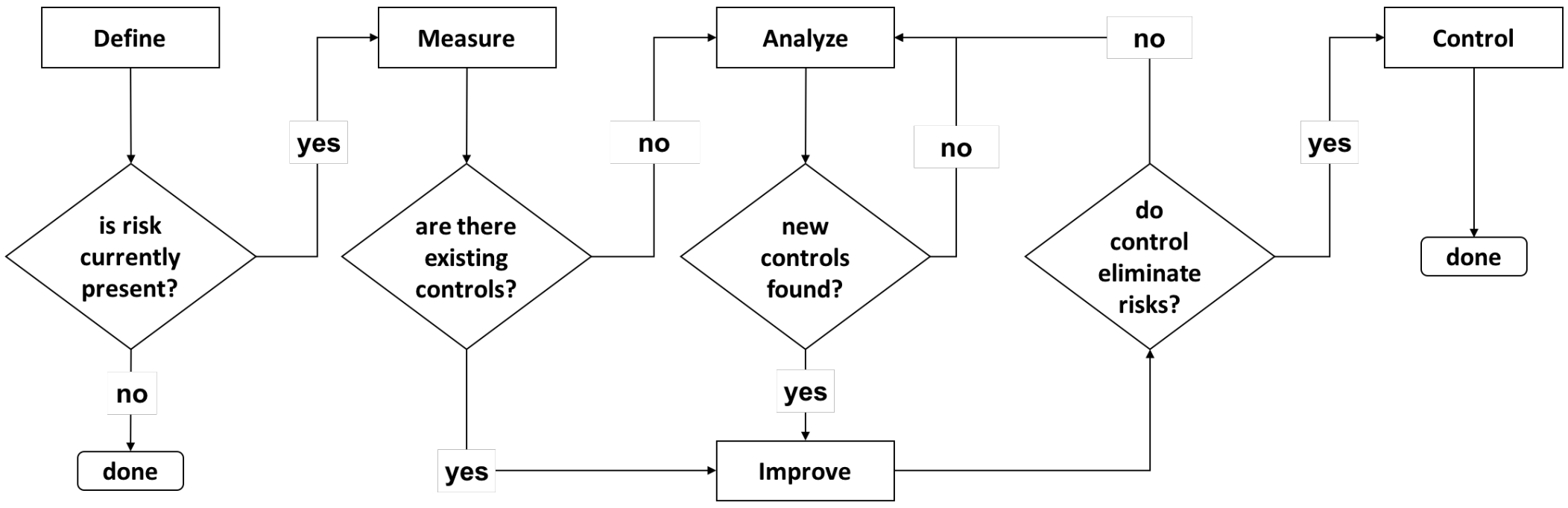

Figure 6. High-Level DMAIC Process Flow. (adapted from Hambleton, 2007)

As previously mentioned the proposed process in Figure 6 makes use of the DMAIC steps which are adapted to serve the purpose of this paper. The entire process, which is designed to take care of security risks from identifying them to preparing a solution including a cost measurement, is a never-ending process. This process should be executed at regular intervals to guarantee that security risks are identified as early as possible. The following list describes the proposed process steps in detail:

*1) Define:* the process starts by identifying and defining security risks for a cloud service. This can be achieved either by implementing a monitoring system [25] or by using any other method to identify security issues. In our use case the OWASP top 10 cloud security risks list was used and categorized by relevance and severity of consequences. Even though this paper only examined the top three risks (R4, R6 and R9) this method can be used to identify all risks as listed by the OWASP.

*2) Measure:* if the risks from the define phase are still present in the current cloud service, the next step would be to categorize them and check whether there were existing security controls to fix the issues. The categorization could be done by a risk analysis as described in Section A and shown in Figure 5. The output of this step is a list of risks grouped by their relevance to the current cloud service / process and the resulting damage, if the issue is not resolved. If there are existing security controls which could be applied the Analyze step can be skipped.

*3) Analyze:* in this step, the main goal is to find solutions to eliminate all identified and categorized security risks which cannot be solved with current security controls. As described in Section B, we used security controls (S17, S10 and S9) as proposed by an established standard (ISO 27017) to solve the top three issues (R4, R6 and R9) identified and categorized in the previous Section A. The output of the Analyze phase is a list of security controls, which can be applied to eliminate the identified issues.

*4) Improve:* so far, we found out WHAT the security risks are, IN WHAT ORDER they need to be solved and WHAT controls could be used to do so. Based on that, the next step is to create a security control implementation plan and to verify whether the control eliminates the risks or not. The plan includes the implementation steps and a list of all resources (i.e.: all expenses like personnel, software, hardware, etc.) required. Next, these steps are prioritized and tested to see if the solutions resolve the problem.

*5) Control:* in the final step a comparative cost analysis shows how much it would cost to implement the security controls (S17, S10 and S9) to eliminate the risks (R4, R6 and R9). In the following cost analysis software-, hardware-, process-, and operational-factors are considered to calculate the implementation expenses. These factors are based on the presented use cases, the resulting security risks and security controls. Software- and hardware-costs are represented by monetary costs, while process- and operational-costs are measured by man-days (MD). In this regard, an MD is defined as an eight-hour day with an assumed hourly rate of EUR 100. This simplified cost metric will be enhanced in future work by investigating more generic cost metrics. The following table shows the cost analysis measuring the resulting implementation costs:

TABLE II. COST ANALYSIS

| Costs | Security Controls / Risks | | |
|---|---|---|---|
| | ***S17 / R4*** | ***S10 / R6*** | ***S9 / R9*** |
| Software, Cloud Service | 17 €[a] | 17 €[a] + 230 €[d] | 17 €[a] + 230 €[d] |
| Hardware | 360 €[b] | - | 90 €[e] + 570 €[f] |
| Process | 400 €[c] | 800 €[c] | 800 €[c] |
| Operation | 400 €[c] | 800 €[c] | 800 €[c] |
| **monthly** | 1.177 € | 1.847 € | 2.507 € |

a. Amazon Lambda Pricing. (AWS: Development & Compute Time)

b. Amazon Echo Pricing. (3 x Amazon Echo + 3 x Amazon Echo reserves)

c. Process-/Operation-Costs. (1 MD = 8 hours * 100 € = 800 €)

d. Voice Biometrics. (VoiceIt: 10.000 API Calls per Month)

e. Kensington Desktop Lock for Amazon Echo. (3 x Desktop Lock)

f. Motorola G5 with Finger Scan Sensor. (3 x Mobile Phone)

## V. CONCLUSION AND FUTURE WORK

In this paper, we introduced a use case of for smart business by using smart devices and cloud services. We explored the technical specifications and services of Amazon Echo and explained the ASK. In this regard, we pointed out that the built-in skills of Alexa could be extended by custom built skills. Furthermore, we have proposed an architectural design to illustrate two points. First, to show how simple it is today to use smart devices and cloud services to build powerful smart businesses with little effort. Second, we have illustrated that IoT can have a big impact both on your business and on security. In the end of section III we presented two use cases showing how Amazon Echo and Alexa can be combined to enable direct, automated and voice-based communication to automate routine tasks.

Finally, in Section IV we proposed a high-level DMAIC process to identify and categorize security risks, analyse and prepare security controls, plan their implementation and in the end, measure the cost involved when implementing them. In this regard, we first presented a method to identify (Define: OWASP) the top ten security risks and label them by their relevance to the use cases and the impact they would have, if the risks were not eliminated (Measure: Risk Analysis). Next, we presented an established standard (ISO 27017) which provides security controls for securing cloud services. Then, we presented solutions according to ISO 27017 to the top 3 security risks from the risk analysis (Analyze). In addition to that we presented how these security controls could be implemented and what resources would be needed to do so (Improve). In the last step, we evaluated the resulting expenses when applying the presented security controls to eliminated the analysed security risks (Control: Cost Analysis).

In summary, we explored the possibilities of smart devices and cloud services to enable smart businesses. In addition to that we proposed methods to identify and categorize security risks, to analyze solutions, plan implementation steps and to measure the costs to do so. The main contribution of this paper is the initial investigation on an approach for how to measure the cost of security in smart business by a combination of the above methods in a simple process. This will be enhanced in future work by considering multiple other use cases and hence more security requirements including a more generic metric for cost calculation. We will furthermore also investigate alternatives to the OWASP, the ISO standard and Six Sigma to verify a broader application of our work.

## VI. ACKNOWLEDGEMENT

Research leading to these results has received funding from the EU ECSEL Joint Undertaking under grant agreement n° 737459 (project Productive4.0) and from the partners national funding authorities FFG on behalf of the Federal Ministry for Transport, Innovation and Technology (bmvit)and the Federal Ministry of Science, Research and Economy (BMWFW).